\documentclass[12pt]{article}

\usepackage[utf8]{inputenc}
\usepackage[T1]{fontenc}
\usepackage[english]{babel}
\usepackage{bold-extra}
\usepackage{amsmath,
	    amsfonts,
	    amstext,
	    amssymb,
	    mathrsfs,
	    mathpartir,
	    stmaryrd,
	    latexsym,
	    enumitem,
	    proof,
	    graphicx,
	    color,
	    hyperref,
	    comment,
	    mathtools}

\usepackage{xcolor}

\usepackage[normalem]{ulem}

\usepackage{geometry}
\usepackage{setspace}
\onehalfspace

\newcommand{\ms}[1]
	{\null\ifmmode\mathord{\mathcode`-="702D\it #1\mathcode`\-="2200}
	\else$\mathord{\mathcode`-="702D\it #1\mathcode`\-="2200}$\fi}

\newcommand{\cws}[2]
	{\\ \centerline{$#2$} \\[-#1pt]}

\newcommand{\bibtrick}[1]
	{}

\newcommand{\auxarrow}
	{\mathop{\longrightarrow}}

\newcommand{\wauxarrow}
	{\mathop{\Longrightarrow}}

\newcommand{\eqdef}
	{\triangleq}

\newcommand{\pco}[1]
	{\mathop{\Vert_{#1}}}

\begin{document}

\title{\textbf{On the Operational Resilience of CBDC:
	       Threats and Prospects of \\
	       Formal Validation for Offline Payments}}

\author{\hspace*{-0.3cm} Marco Bernardo$^{1}$ \!
	Federico Calandra$^{1}$ \!
	Andrea Esposito$^{1}$ \!
	Francesco Fabris$^{2}$ \\[0.2cm]
	\hspace*{-0.4cm} $^{1}$Dipartimento di Scienze Pure e Applicate, Univ.\ di Urbino \\
	\hspace*{-0.25cm} $^{2}$Dipartimento di Matematica, Informatica e Geoscienze, Univ.\ di Trieste}

\date{}

\maketitle


\begin{abstract}
Information and communication technologies are by now employed in most human activities, including economics
and finance. Modern computers have reached an extraordinary power in terms of information processing,
storage, retrieval, and transmission. However, several results of theoretical computer science imply the
impossibility of certifying software quality in general.

With the exception of safety-critical systems, this has primarily concerned information processed by
confined systems, with limited socio-economic consequences. \linebreak In the emerging era of technologies
for exchanging tokenized assets and digital money over the Internet, such as in particular central bank
digital currency (CBDC), even a minor bug could trigger a financial collapse.

Although the aforementioned impossibility results cannot be overcome in an absolute sense, there exist
formal methods that can provide correctness assertions for software system models under suitable conditions.
We advocate their use to \linebreak validate the operational resilience of software infrastructures for CBDC
by framing the offline payment problem, organizing its threat landscape, and outlining a formal methods
methodology, illustrated by a minimal proof of concept, that we argue should underpin CBDC design and
deployment.
\end{abstract}

\newpage

\section{CBDC and Digital Euro: A Digital Form of Cash}
\label{sec:cbdc_digital_euro}

Central bank digital currency (CBDC) is set to become the latest technological step in the long history of
money. After barter, cowrie shells, metal coins, and banknotes, the 20th century saw the introduction of
digital money through the immateriality of bits stored on computers operated by banks, with account balances
managed via bank transfers and debit or credit card transactions. In our modern developed societies, where
most of our activities are digitally mapped, the way people pay has a digital dimension too, \linebreak with
more and more consumers around the world increasingly preferring to pay by using digital electronic devices
rather than traditional banknotes or coins or even physical cards. Therefore, in a context where the use of
physical cash is declining, the idea of a CBDC directly controlled by the central bank has gained growing
attention. It would represent the next step forward for legal tender issued by the central bank,
guaranteeing monetary sovereignty and financial stability in the presence of private digital money pegged to
foreign currencies.

A CBDC would be a digital form of cash that complements banknotes and coins, giving people an additional
choice of payment modalities. It would be issued by the central bank and freely available to all citizens.
If managed under a federalist model, commercial banks would play a key role as the front-end interface with
customers. Under this paradigm, commercial banks would remain responsible for managing the relationships
with end customers, activating the necessary know-your-customer/anti-money-laundering (KYC/AML) procedures,
distributing approved CBDC wallets to customers, and resolving any issue that might arise. On the other
hand, citizens would be free to receive their salaries, pensions, and other types of small incoming payments
directly into their own wallets. At the same time, they would be able to spend their money by using online
or offline services provided by the central bank, which would be secure and guarantee the appropriate level
of privacy.

It is well known that today 137 countries and currency unions, representing 98\% of the global gross
domestic product, are exploring CBDC solutions through ongoing investigations, studies, models, and pilot
projects on the topic~\cite{AC_2025}.

In particular, the European Central Bank (ECB) planned to introduce digital euro~\cite{ECB_2025} as its
CBDC. It will be freely available and free of charge for any digital payment to all citizens in the euro
area. After an Investigation Phase that took place from October 2021 to October 2023, aimed at assessing
feasibility, benefits, design options, possible risks, and issues, a two-year Preparation Phase was launched
in November 2023. The objective was to test the potential issuance, define the necessary rules, identify the
infrastructure required for smooth and secure usage, and establish the legal frameworks and operational
rulebook under which to pursue the issuance. Another key element referred to conducting user research to
ensure that the digital euro project under development is able to address the vast majority of the needs of
European citizens.

Since physical cash enables payments in the absence of electricity and network coverage, in the design of
the digital euro special attention should be devoted to offline payments. In addition to guaranteeing the
financial inclusion of unbanked individuals, this solution is also intended to keep the digital euro
operational during emergencies such as power or network outages caused by infrastructure failures, natural
disasters, cyberattacks, geopolitical conflicts, or other similar events. In any case, the ECB remains
committed to ensuring that cash continues to be accepted everywhere in the euro area. It is worth noting
that, as of today, there is no single European digital payment solution capable of covering the entire euro
area: 13 out of 20 countries rely on non-EU credit card issuers for card payments. In contrast, the digital
euro will be a fully European digital form of payment, freely accessible and universally accepted across all
euro area countries. From the legal point of view, in 2023 the European Commission approved a proposal for
the regulation of the digital euro~\cite{ECB_2025}.

In this paper we lay the groundwork for the systematic use of formal methods in the validation of CBDC, with
special emphasis on offline payments. Specifically, we make the following contributions:

	\begin{itemize}

\item We motivate operational resilience as a first-class, socio-economically critical requirement for CBDC
infrastructures deployed on the open Internet (Sections~\ref{sec:cbdc_resilience} and~\ref{sec:sw_quality}),
recalling the theoretical limits of software assurance and formal methods for their mitigation.

\item We consolidate, from central bank sources, a structured taxonomy of threats and vulnerabilities
specific to offline CBDC payments (Section~\ref{sec:offline_payments}), including the open problem of
devices that remain offline for extended periods.

\item We propose a three-phase formal-methods-based methodology for validating CBDC infrastructures, which
we instantiate on a proof of concept (Section~\ref{sec:formal_methods}).

	\end{itemize}

\noindent
We then conclude with some final remarks (Section~\ref{sec:concl}).

\section{The Challenge of Operational Resilience in CBDC}
\label{sec:cbdc_resilience}

Information and communication technology has become increasingly pervasive due to its daily application to
basically any human activity, including economics and finance. After the invention of printing in 1400 --
which fostered a higher diffusion of knowledge -- and the industrial revolution in 1700 -- which widened
human physical capabilities through the introduction of mechanical machines able to carry out automated
processes -- the digital transformation, started in mid 1900 with the advent of electronics, is continuously
extending human cognitive capabilities through programmable devices like computers and smartphones and the
rapid worldwide propagation of digital data through the Internet.

Computing systems, along with software methodologies and programming languages needed to design and
implement them, have faced a growing complexity. From sequential systems, in which a single operation at a
time can be executed, the first evolution has been several decades ago toward concurrent and distributed
systems~\cite{Mul93}, in which multiple computing devices run simultaneously and exchange information
through, respectively, a shared memory or message passing. The second evolution, started with the financial
crisis of 2008 and the consequent decreasing trust in institutional intermediaries, has led to decentralized
systems~\cite{Nak08}, where parties not knowing each other perform transactions by relying on a distributed
ledger and a consensus mechanism that ensure trust in an algorithmic way.

Several advanced paradigms are by now in place. They range from mobile computing, where devices are no
longer tied to physical locations (IoT -- Internet of Things~\cite{Mil20}), global computing, which
implements the abstraction of a single computer accessible anywhere anytime (cloud, edge, fog
infrastructures~\cite{EB23,BN19}), and autonomic computing, typical of systems capable of adapting to
unpredictable changes (via sensors, actuators, policies~\cite{LMD13}), to machine learning, deep learning,
and large language models (AI -- Artificial Intelligence~\cite{RN20,WBK20,KKSS24}). Moreover, ongoing
research is focusing on future paradigms such as quantum computing~\cite{NC10}, which should make
computation faster, and reversible computing~\cite{Lan61}, which should reduce energy consumption.

The increasingly complex computer and network infrastructures in use nowadays have profoundly transformed
the backbone of all information processes required to sustain the full functionality of the intricate web of
interactions among individuals, institutions, and digital systems. Some of these interactions are very
critical; think, e.g., of the software used to regulate the flight traffic of an international airport or
the computer procedures necessary to control the core of a nuclear power plant. Digital financial
transactions should be considered critical as well, because of the dramatic socio-economic impact of
possible errors in their enabling infrastructures.

This high complexity exacerbates the issue of software quality assurance; the interested reader is referred
to~\cite{WikiSwBugs} for a list of major software disasters and their consequences in terms of costs, human
lives, and environmental impact. Software quality is not only about code functional correctness, because it
is equally important to avoid poor performance, security breaches, and bad usability. Moreover, the normally
adopted approach of testing software is not enough, as it does not guarantee the absence of errors.
Therefore, it is of paramount importance to adopt software verification techniques, which are mostly based
on the early development of software models and hence need to be accompanied by model-driven software
engineering methodologies.

While the presence of errors in a non-safety-critical software designed to handle information may be merely
annoying or pose some unpleasant consequences, an infinite loop, a logical flaw, or a structural weakness in
a system designed to manage digital money or tokenized assets can lead to devastating outcomes from a
socio-economic viewpoint. Consider the scenario in which a hacker is able to create CBDC tokens out of thin
air, steal monetary reserves from a bank, or generate, steal, or forge smart contracts associated with
tokenized assets. In such cases, the capitalization of an entire financial asset could collapse within hours
or even minutes, as investors trust -- the foundational requirement for a healthy financial system -- would
be irreparably damaged. Furthermore, if we are dealing with a CBDC, the consequences could trigger
instability across the entire financial system and potentially take on a geopolitical dimension.

The trend of cybercrimes involving hackers stealing money is closely monitored by specialized institutions.
For example, the International Monetary Fund (IMF) reports more than 20,000 cyberincidents in the
traditional financial sector between 2004 and 2023, with total losses of 10 billion USD (see Figure~3.2
of~\cite{IMF_2024}). In cryptocurrencies and the decentralized finance (DeFi) sector, both supported by
public blockchains, Chainalysis reports nearly 14 billion USD in losses from 1,329 hacking events between
2015 and~2024~\cite{Chainalysis_2025}. Even if not all the breaches described above are necessarily linked
to structural flaws in the software, the phenomenon remains deeply concerning.

The reader should note a fundamental difference between the infrastructure of the traditional financial
sector and that of public cryptocurrencies and the DeFi sector. In the former, the networks connecting
financial institutions are private and not directly accessible via the Internet. One example is the SWIFT
interbanking system, which has nevertheless been the target of significant attacks such as Banco del Austro
(Ecuador) in January 2015, Bangladesh Bank in February 2016, and Banco de Chile in May 2018. In contrast,
the entire cryptoasset sector -- whose market capitalization now exceeds 3~trillion USD -- is directly
accessible via the Internet and therefore more exposed. The CBDCs that are being introduced, as well as
those already in operation such as the Chinese e-CNY, are deployed with Internet connectivity to support
peer-to-peer transactions between users or users and institutions, interoperability with wallets, commercial
banks, and payment systems, real-time updates, centralized control, and, last but not least, financial
inclusion. This implies that exposure to hacking threats will significantly affect current and future CBDCs.

\section{Formal Methods for Ensuring Software Quality}
\label{sec:sw_quality}

Theoretical computer science provides several impossibility results, which arise from the undecidability of
many computational problems. In the specific case of software quality, suppose that we create a program $P$
to compute a function $f_{P}$. If $P$ is not properly designed, its execution may be subject to two
different kinds of errors. The first one corresponds to a situation in which the program enters an infinite
loop; when this happens, the only feasible action is to interrupt the computation, that is, to shut down the
system. The second one, if possible, is even worse: the system appears to be working correctly, but the
function $f_{W}$ it is computing is actually a wrong one, i.e., $f_{W} \neq f_{P}$.

To overcome these two highly undesirable situations, one might consider building two test programs.  The
first one, $T_{l}$, is designed so that, when supplied with the code lines of $P$, it can answer ``Yes'' or
``No'' in finite time to the question: ``Does $P$ enter an infinite loop in some situations?''. The second
one, $T_{f}$, is able to answer ``Yes'' or ``No'' in finite time to the question: ``Does $P$ compute the
function $f_{P}$?''. Unfortunately, two fundamental results of computability theory -- due to
Turing~\cite{Tur36} and Rice~\cite{Ric53} -- establish the impossibility of constructing such tests for all
programs. The practical consequence of these two results is that upon every software release there remains a
residual risk \linebreak of malfunction.

In the last 50 years, several approaches have been developed to assess the quality of software under
conditions that make such a problem decidable. These are collectively called formal methods in computer
science. They are mathematically rigorous techniques for modeling and verifying computing systems, based on
the idea that appropriate mathematical analysis can contribute to a dependable software design similar to
other engineering disciplines. In the following, we recall some of the prominent formal methods.

On the modeling side, formal descriptions of software can be provided at different abstraction levels and
mostly revolve around automata, algebras, and logics.

An automaton~\cite{HMU06} is a graph consisting of a set of states and a set of transitions between states.
Every state of an automaton represents a state of the computation, like for instance the current statement
to be executed in a program along with the value contained in every program variable, while a transition
describes a state change, which can be accompanied by possible inputs and outputs. There is a hierarchy of
automata classes, ranging from finite-state automata to Turing machines, which correspond to the hierarchy
of formal languages established by Chomsky~\cite{Cho59}. Moreover, in addition to traditional automata in
which states are global, there are variants called Petri nets~\cite{Pet62,Rei85} in which the underlying
graph is bipartite; in this way, states can be represented in a distributed fashion across loci of
computations, which is well suited for concurrent software.

Process algebras~\cite{Mil89a,Hoa85,Hen88a,BW90,BPS01} are description languages that provide a number of
operators whereby to obtain complex system models by combining simpler ones. Among these operators we
mention sequential, alternative, and parallel compositions, which allow one to express the fact that two
processes -- intended as the behaviors of two systems -- are respectively executed one after the other,
alternative to each other, or run in parallel. \linebreak A fundamental notion in this setting is that of
behavioral equivalence, with bisimilarity -- which expresses the capability of mimicking each other's
behavior stepwise -- being one of the most important approaches. The objective is to identify syntactically
different processes that feature the same observable behavior. Compositional modeling is accompanied by
compositional reasoning when the considered equivalence is a congruence -- \linebreak i.e., it allows to
replace equals with equals inside arbitrarily complex process terms -- with respect to the process algebraic
operators.

Logics are more abstract than automata and process algebras, in that they express the behavioral properties
that a computing system should possess. The simplest example is given by Hoare logic~\cite{Hoa69,Dij75} and
its variants, including separation logic~\cite{Ohe19}, which are used to enrich programs with annotations
containing logical formulas that state properties that should be valid in those points of programs
execution. Other examples are modal and temporal logics~\cite{HC77,GHR94}. They are extensions of classical
logic respectively including modal or temporal operators. The former denote the possibility or the necessity
of performing certain activities in a given order. The latter express a property to be valid in the next
step or the fact that a certain property has to hold until another one becomes satisfied, along at least one
computation or all possible computations.

On the verification side, we mention model checking, equivalence checking, and \linebreak deductive
verification.

A model checker~\cite{Mcm93,CGP99,BK08} takes as inputs a program model, described as an automaton or a
process algebraic term, and a property it should possess, formalized as a modal or temporal logic formula.
Then it communicates whether the program model meets that property or not. In case of failure, it provides
diagnostic information in the form of a program model execution that does not fulfill the property.

An equivalence checker~\cite{BPS01} considers instead a model of the specification of the program and a
model of the implementation of the program, both expressed as automata or process algebraic terms. Then it
checks whether the two models are behaviorally equivalent to each other or not. In case of failure, it
exhibits a distinguishing modal or temporal logic formula, i.e., a formula that is satisfied by only one of
the two models.

In deductive verification~\cite{Hoa69,Dij75,Ohe19,Pau89,BC04,ABBHSU16}, the properties of interest for a
program are expressed in terms of logical formulas formalizing precondition-postcondition pairs or loop
invariants. These are proof obligations that, once they have been discharged, imply program correctness. For
their verification, possibly interactive theorem provers are typically employed.

The aforementioned modeling and verification techniques, originally developed for functional features of
programs, have been subsequently extended to deal with the software architecture design level and
quantitative aspects too~\cite{Han92,Seg95a,ABCDF95,Hil96,Her02,AILS07,DKV09,ABC10,GM14}. In this way, it is
possible to address also systems including probabilistic behaviors -- think, e.g., of cryptocurrencies based
on proof-of-stake consensus -- or real-time constraints -- think, e.g., of the deadline by which a bank
transfer can be canceled. Moreover, they have been adapted to address security properties as well, like for
instance the use of equivalence checking to assess noninterference of information flows across different
security levels~\cite{GM82,FG01,EABR25}. Many software tools have been implemented to support formal
modeling and verification, like for instance CADP~\cite{GLMS13}, mCRL2~\cite{GM14}, Spin~\cite{Hol03},
NuSMV~\cite{CCGR00}, Uppaal~\cite{LPY97}, PRISM~\cite{KNP11}, Modest Toolset~\cite{HHHK13},
GreatSPN~\cite{ABCDF95}, PEPA Workbench~\cite{Hil96}, TwoTowers~\cite{ABC10}, Isabelle~\cite{Pau89},
Rocq~\cite{BC04}.

In this paper, we advocate the use of formal methods to validate the operational resilience of extremely
critical software infrastructures such as those underlying CBDCs, with special emphasis on offline payments
as they constitute a troublesome matter.

\section{Issues and Risks of Offline Payments in a CBDC}
\label{sec:offline_payments}

One of the main innovations in the proposal for a CBDC is the necessity of enabling digital offline
payments. This means that it would be possible to transfer value -- the digital money of the CBDC -- between
the wallets of two physical devices without requiring an Internet connection to any ledger system. This
could be due to the total absence of Internet coverage or temporary or local difficulties in
telecommunications connectivity as a consequence of a system outage. However, the offline solution could
also be chosen as a way to preserve the highest level of privacy during transactions, as this kind of money
transfer would replicate some key features of a cash exchange. This could address the consequences of the
declining use of cash~\cite{BoE_2020}, with the CBDC acting as a medium for pseudonymous or anonymous
payments~\cite{EUBOF_2021}.

Another important advantage of the offline payments solution is that it would enable the financial inclusion
of unbanked people, who could use offline payments as a substitute for cash. In a scenario in which cash is
gradually abandoned as a medium for transferring value, commercial banks might find it economically
disadvantageous to offer services to unbanked people, who are de facto financially excluded groups.
Therefore, the introduction of offline payments as an instrument of financial inclusion is becoming a
necessity to support the most vulnerable parts of our societies, which would otherwise be excluded from the
economy. To achieve this, only a minimal technological infrastructure would be required: a
smartphone, a basic computer, or a laptop, with proximity connection.

In a report by the World Bank Group, it is estimated that globally 1.4 billion adults are
unbanked~\cite{WBG_2021}. As for the EU, a 2024 Economic Bulletin of the ECB~\cite{ECB_2024} states that
``\emph{nearly one out of five adults (19.4\%) in the euro area reports not having either debit or credit
cards or payment accounts}''. As a consequence, the offline payments solution with cash-like features is
strongly recommended as an instrument for financial inclusion also in the context of the future digital
euro. Indeed, it constitutes the second requirement in the ``Report on a digital euro'' by the
ECB~\cite{ECB_2020}:

	\begin{quote}
\small
\textbf{Requirement 2 (R2): cash-like features.} To match the key distinctive features of cash, a digital
euro aiming to tackle a decline in the acceptance of cash should permit offline payments. Moreover, a
digital euro should be easy for vulnerable groups to use, free of charge for basic use by payers and should
protect privacy. \linebreak It should have a strong European branding.
	\end{quote}

A further element supporting the offline payments solution derives from a survey conducted by the Innovation
Hub of the Bank for International Settlements (BIS). \linebreak It shows that ``\emph{49\% of central banks
surveyed consider offline payments with retail CBDC to be vital, while another 49\% deemed it to be
advantageous}''~\cite{BIS_2023_Polaris_1}.

Unfortunately, even if highly recommended, the offline payments solution presents some challenges and issues
from operational and technical points of view. In an expository study by the Bank of Finland, a case study
named ``Project Pluto'' is discussed, aimed at investigating the risks of offline CBDCs~\cite{BoF_2023}. In
contrast to the Riksbank's e-krona pilot in
Sweden~\cite{BoS_2017_Ekrona_1,BoS_2018_Ekrona_2,BoS_2022_Ekrona_3}, the Pluto application focuses on the
token-based model and relies on reusable tokens.

In a note of the Bank of Canada~\cite{BoC_2023}, it is explicitly written that:

	\begin{quote}
\small 
From a financial risk perspective, concerns exist that an extended offline solution may become a target for
fraud and financial crime. These concerns, in addition to security concerns, mean that extended offline
functionality implies some risk.
	\end{quote}

The Bank of England also aimed to assess whether it is technically feasible to implement a secure offline
payment functionality for a digital pound. The conclusions highlight some critical issues~\cite{BoE_2020}:

	\begin{quote}
\small 
This project demonstrated that while it might be technically feasible to implement an offline payment
functionality for a digital pound, there are trade-offs, particularly around user experience and preventing
double spending and counterfeiting, that make implementing it challenging.
	\end{quote}

But the most detailed study on the secure feasibility of the offline payments solution is the one conducted
by BIS, which has recently produced a series of analytical documents to illustrate ``Project Polaris''. The
first report, entitled ``A Handbook for Offline Payments with CBDC''~\cite{BIS_2023_Polaris_1}, outlines the
current technical solutions for offline payments with a CBDC, the possible risks, threats, and
vulnerabilities associated with them, and the countermeasures to mitigate these issues. The second report,
``A Security and Resilience Framework for CBDC Systems''~\cite{BIS_2023_Polaris_2}, is devoted to a study of
the objectives and design criteria required to achieve security and resilience in a CBDC framework. The
third report, ``Closing the CBDC Cyber Threat Modeling Gaps''~\cite{BIS_2023_Polaris_3}, addresses the
problem of analyzing the adequacy of current cybersecurity standards and frameworks through real-world
observations of security flaws; the goal is to develop cyber threat models that need to be addressed to
achieve full control over the security necessary in a CBDC framework. The fourth and final report, entitled
``A High-Level Design Guide for Offline Payments with CBDC''~\cite{BIS_2023_Polaris_4}, outlines the
suggested design choices for offline payment solutions, also analyzing the security, risk, and impact on
resilience of the offline system.

From all these studies, we can deduce many of the main issues and vulnerabilities of offline payments that
need to be addressed.

Generally speaking, as noted in~\cite{BZKWCC_2024} we must state that basing part of the security of offline
solutions on the assumption of inviolability of secure elements\footnote{A secure element is a
microprocessor chip with enhanced security protections that should prevent unauthorized access. It can be
embedded in smart cards and smartphones or in removable subscriber identity module (SIM) cards.} is simply
naive. Furthermore, in general even a single compromised device could print and spend counterfeit money
without restrictions. For example, a hacker able to break a single secure element could send counterfeit
money to a colluding device. Moreover, without adequate countermeasures, offline payment recipients cannot
distinguish legitimate payments from counterfeit ones, so central banks would be forced to bear these risks
of fraud.

We briefly illustrate here a non-exhaustive tentative list of threats based in part on the first BIS report
of ``Project Polaris''~\cite{BIS_2023_Polaris_1} and on the expository study of the Bank
of~Finland~\cite{BoF_2023}:

	\begin{description}

\item[Counterfeiting by taking control of a device.] If a payment device is allowed to generate and redeem
tokens, a bad actor could take over the device and generate additional tokens.

\item[Counterfeiting by a physical breach.] Using an offline CBDC payment feature means that there must be
some digital representation -- a binary string -- of the CBDC tokens stored inside a physical device, be it
a smart card, a smartphone, a hardware wallet, or other. A bad actor could try to clone or manipulate this
string through a physical attack on the device, in order to (double-)spend the balance with another device
or alter the original balance. Note that the attacker could be in possession of the device, with unlimited
time to mount such attacks without being detected.

\item[Counterfeiting by a cryptanalysis breach.] This is a non-intrusive breach such that the security of a
transaction -- which is based on cryptographic protocols -- is compromised by the knowledge of the private
key, which is stolen from the owner or derived through a successful cryptanalytic attack.

\item[Side-channel attack.] The physical devices used in the framework of offline payments are digital
electronic devices that can leak electromagnetic fields while operating. It is therefore possible, for a bad
actor, to analyze the variations of these fields and perform signal analysis on the electromagnetic
radiation. This could allow them to deduce some information about the electrical quantities that physically
represent the bit strings associated with the digital representation of the CBDC tokens or the~private key
involved in a transaction.

\item[Third-party device compromise.] The physical devices used to perform offline payments are third-party
devices not under the direct control of the central bank. They might be affected by structural hardware or
software weaknesses or flaws, which could be exploited by a bad actor to obtain control of the cryptographic
keys or directly of the digital representation of the CBDC tokens.

\item[Third-party device complexity.] Offline CBDC payment systems may be forced to use technologies that
are not well understood by the internal staff of a central bank. \linebreak This could increase reliance on
third-party expertise and clearly constitutes an operational risk.

\item[Device obsolescence.] As in our common experience, software and hardware devices need to be updated
when necessary to ensure vendor support. Deprecated software versions or unsupported hardware can offer a
trapdoor to bad actors, who can exploit bugs or other vulnerabilities no longer fixed by vendors to breach
the system.

\item[Double spending.] The double-spending issue can be associated with several previously described
situations or contexts. Moreover, note that the double-spending issue for offline CBDC payments is different
from the more famous problem of double-spending cryptocurrencies on a public and disintermediated
blockchain. In the latter case, the problem of consensus on an asynchronous network must be solved, but it
is well known that the asynchronous version of the Byzantine generals problem does not admit solution in
theory~\cite{FLP85}. Nevertheless, practical solutions based on proof-of-work~\cite{Nak08} and
proof-of-stake~\cite{KN12} protocols offer some level of security. \linebreak In contrast, in the case of
offline payments for a CBDC, trust needs to be ensured by the tamper-resistant nature of the user device and
the shared cryptographic protocols of the wallet. It is therefore necessary to guarantee that
double-spending attempts will be rejected by the payee's wallet.

\item[Fraud.] Since bad actors may attempt to persuade users to pay them by impersonating payees known to
the user, it is necessary to design an offline system with the capacity of uniquely identifying payees to
prevent this kind of fraud. This conflicts with requirements that support privacy by design.

\item[Lost value.] The digital representation of the CBDC tokens stored inside a physical device poses
another kind of risk, because the value associated with it can be lost under several scenarios. Examples are
the device being lost or broken, the user becoming unable to use the device or losing the credentials to use
it, or the transaction being torn, in the sense that the tokens have left the payer's wallet but have not
been received by the payee's wallet due to an interrupted transaction.

\item[Insider threat.] Internal staff of the central bank with roles requiring privileged access to
technological infrastructures, such as IT administrators or system operators, could accidentally or
deliberately affect the regular system operativity, acquiring knowledge aimed at tampering with the system,
stealing funds, changing balances, or other malicious actions.

\item[Scalability of risk in counterfeiting.] If we compare the effects of counterfeiting between the old
technology of physical cash based on banknotes and the new technology of an offline payment system based on
a CBDC, a striking difference in the capacity of a bad actor to scale counterfeiting becomes evident.
Counterfeiting banknotes is a production-heavy process because, apart from the technical expertise and
specific skills that are required, the bad actor also needs special inks, holograms, special printing
machinery, and raw materials. In other words, there exist several production-related factors that limit the
ability to scale up the counterfeiting of banknotes. On the contrary, counterfeiting digital tokens scales
very efficiently. Once the offline CBDC system has been compromised, due to the immaterial nature of the
digital representation of the CBDC tokens it is virtually possible to generate an unlimited number of tokens
at a cost that is essentially zero, as the only physical resources necessary to perform the counterfeiting
are a computer and/or a smartphone. This implies that the consequences in the loss of trust in an offline
CBDC system would be far more severe than in the case of banknotes. Moreover, since in general it is the
payee who bears the risk of economic loss in case of receiving a counterfeit payment, the attractiveness of
offline payments based on the CBDC could decline considerably. While for a banknote a payee has some
capacity to check and verify its authenticity -- e.g., by using technological devices dedicated to this task
-- in the case of an offline payment the payee has no real instrument for checking the authenticity of the
transaction.

	\begin{figure}[t]
\centering
{\includegraphics[width=0.8\textwidth]{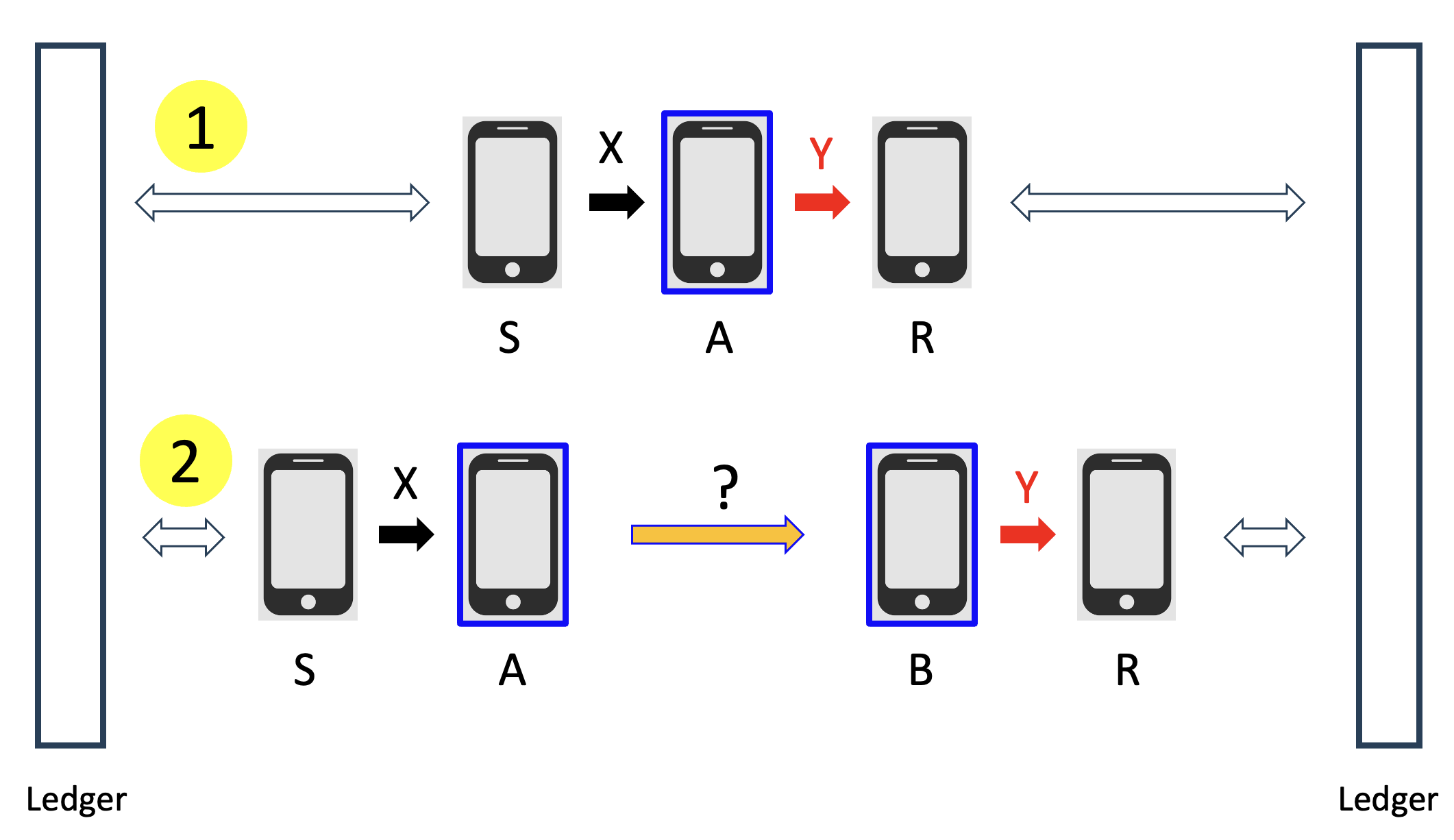}
\caption{The risk of devices staying offline: the bordered ones always remain offline~\cite{BoF_2023}.}
\label{fig:offline_devices}}
	\end{figure}

\item[The risk of devices staying offline.] This is probably the most insidious problem that, to the best of
our knowledge, has not been solved yet. Suppose that there are two chains of offline transactions, \#1 and
\#2, as in Figure~\ref{fig:offline_devices}. In both chains, an amount X of tokens is moved, while the
bordered smartphones always remain offline~\cite{BoF_2023}. \\
In transaction chain \#1 the sender S is in offline mode and transfers X tokens to~A. Then A transfers Y
counterfeit tokens to the final receiver R. When both smartphones S and R switch to online mode and connect
to the ledger, we assume that it is possible to realize that Y tokens are counterfeit. At this point, using
the transaction history of the devices, the system can plausibly presume that A was the fraudster even if A
stays offline. This is because, in principle, it is possible to mirror the transaction chain through
information received from wallets S and R when they enter online mode again. \\
The situation is completely different in transaction chain \#2. When S and R come back online after the
offline transactions, the system can certify that S has sent X tokens to A and R has received Y counterfeit
tokens from B. But it is impossible to detect whether A or B has produced the counterfeiting. Likewise, it
is also impossible to deduce whether there have been more devices in between A and B.

	\end{description}

All the risks, issues, and vulnerabilities described above need to be tackled under a global approach able
to formally validate, from the security and anti-counterfeiting point of view, all the concurrent processes
in action when the offline payment system associated with a CBDC is activated. It is well evident that a
loss of trust in the financial institutions behind a CBDC, represented by central banks, could result in a
deadly blow to the entire financial system.

\section{A Methodology Based on Formal Methods}
\label{sec:formal_methods}

The formal methods and related tools mentioned in Section~\ref{sec:sw_quality} have been successfully used
in a number of case studies, some of which conducted in collaboration with industries. They cover
distributed algorithms and systems, communication and coordination protocols, telephony systems, mobile
agents, operating systems, database management, robotics, hardware/software codesign, embedded software,
security and cryptography, software architectural styles, bioinformatics, healthcare, power management,
flood control, railways, avionics, space mission control, autonomous vehicles, multimedia, games, Internet
of things, cloud computing, web services, human-computer interaction, e-government, manufacturing systems,
and business systems.

As far as economics and finance are concerned, the application of formal methods is rather limited. In the
case of traditional centralized systems, we are aware of very few case studies; among them we
mention~\cite{And_1999,LRK_2013,SIR_2013,PI_2017,MMNOSV_2018,SS_2019,Sto_2023}. In the case of modern
decentralized systems, apart from some works about blockchain consensus
like~\cite{VYC_2022,Ves_2023,BEFRG_2025}, the formal methods literature is focusing on smart
contracts~\cite{Sza94}; see, e.g., \cite{BZ_2018}.

As for CBDC and offline payments, to the best of our knowledge there are no applications of formal methods.
We thus complete this section by showing how they could be used in this critical setting.

\subsection{Formal Model of the CBDC}
\label{sec:cbdc_modeling}

The first phase of our methodology is to develop a formal model of the CBDC of interest, like the digital
euro, together with the related online and offline services. The model should take into account the most
important functional, performance, security, and usability aspects of the entire system. Among the various
formalisms presented in Section~\ref{sec:sw_quality}, we expect that process algebras may play a key role
due to their compositional nature. This inherently supports the interplay of a multitude of submodels for
the various system components such as the central bank, commercial banks, digital ledgers, user wallets, and
online and offline transactions.

A prerequisite for this phase is the establishment of a fruitful collaboration between all stakeholders, in
particular the central bank and formal methods researchers. Without knowing the specification defining a
system of interest, it is not possible to build any model for that system.

\subsection{Verification of the Formal Model}
\label{sec:cbdc_verification}

The second phase focuses on the verification of the model that has been built in the first phase. We expect
to employ both model and equivalence checkings. In the case of the former, the idea is to formalize the
properties of interest via suitable modal or temporal logic formulas, which should account for all major
correctness, efficiency, and privacy criteria to satisfy. As for the latter, the appropriate functional,
performance, and secure behaviors have to be described through specific models that are viewed as formal
specifications, then the overall model -- intended to be an implementation that should conform to a
specification -- is checked for equivalence with each of the specific models after possibly hiding some
irrelevant details in the overall model.

\subsection{Model-Based Prototype}
\label{sec:cbdc_prototype}

Once the verification phase is over, the prototyping of the CBDC system with its online and offline services
can start. This should be carried out in a model-driven manner, so that the resulting software preserves by
construction the properties that have been formally proven on its model.

\subsection{A Simple Example}
\label{sec:example}

To illustrate our methodology in the case that process algebras and equivalence checking are employed, let
us consider the design of a CBDC wallet balance management system. In general, this system is composed of an
intermediary bank, a user wallet, and the operations whereby the bank loads CBDC value into the wallet and
the user later spends that value through offline payments. This interpretation is coherent with current CBDC
proposals in which banks and other supervised intermediaries remain the front-end interface with customers,
while offline functionality is meant to support cash-like payments in the absence of network connectivity.
For simplicity, we consider a scenario in which there are a single bank and a single user and the offline
wallet has only two positions, each capable of storing one unit of spendable CBDC value. We also assume that
the units are all identical, so that their specific origin or serial identity is not important.

Since the only observable activities are wallet funding and wallet spending, the specification of the wallet
balance management system -- with which every correct implementation should comply -- can be formalized
through the following process algebraic equations:
\cws{0}{\begin{array}{rcl}
\mathit{WalBal}_{0/2} & \eqdef & \mathit{fund} \, . \, \mathit{WalBal}_{1/2} \\
\mathit{WalBal}_{1/2} & \eqdef & \mathit{fund} \, . \, \mathit{WalBal}_{2/2} + \mathit{pay} \, . \,
\mathit{WalBal}_{0/2} \\
\mathit{WalBal}_{2/2} & \eqdef & \mathit{pay} \, . \, \mathit{WalBal}_{1/2} \\
\end{array}}
where $\mathit{WalBal}_{0/2}$ represents the initial state of the system (in which the offline wallet is
empty), $\mathit{WalBal}_{1/2}$ represents the state in which one position of the wallet is occupied by one
spendable CBDC unit, and $\mathit{WalBal}_{2/2}$ represents the state in which the wallet is full. Operator
$a \, . \, P$ is called action prefix and describes the possibility of executing action $a$ and then
behaving as process~$P$. Operator $P_{1} + P_{2}$ describes a nondeterministic choice between processes
$P_{1}$ and $P_{2}$, which is based on the actions they initially enable.

The underlying automaton -- which can be mechanically obtained from the operational semantic rules of the
process algebraic language -- is the following, where the dashed arrow on top indicates the initial state:
\\[0.2cm]
\centerline{\includegraphics{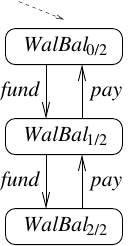}}

There are at least two possible implementations of the wallet balance management system. The first option is
a concurrent implementation, in the sense that the two-position wallet is made out of the parallel
composition of two independent one-position wallet cells:
\cws{0}{\begin{array}{rcl}
\mathit{WB}_{{\rm conc}, 2} & \eqdef & \mathit{Bank} \, \pco{\{ \mathit{fund} \}} \, (\mathit{Cell} \,
\pco{\emptyset} \, \mathit{Cell}) \, \pco{\{ \mathit{pay} \}} \mathit{User} \\
\mathit{Bank} & \eqdef & \mathit{fund} \, . \, \mathit{Bank} \\
\mathit{Cell} & \eqdef & \mathit{fund} \, . \, \mathit{pay} \, . \, \mathit{Cell} \\
\mathit{User} & \eqdef & \mathit{pay} \, . \, \mathit{User} \\
\end{array}}
Operator $P_{1} \pco{S} P_{2}$ describes the fact that processes $P_{1}$ and $P_{2}$ run in parallel and
have to synchronize when executing actions in $S$; in the first equation above this set is empty as the two
wallet cells are independent. Note that $\mathit{Bank}$ (resp.\ $\mathit{User}$) repeatedly tries to fund
with one CBDC unit (resp.\ spend one CBDC unit from) the wallet.

To show that this is a correct implementation of the wallet balance management specification, we have to
investigate the existence of some relation between $\mathit{WB}_{{\rm conc}, 2}$ and
$\mathit{WalBal}_{0/2}$. The first step consists of comparing the automata underlying $\mathit{WB}_{{\rm
conc}, 2}$ and $\mathit{WalBal}_{0/2}$, which are shown below: \\[0.2cm]
\centerline{\includegraphics{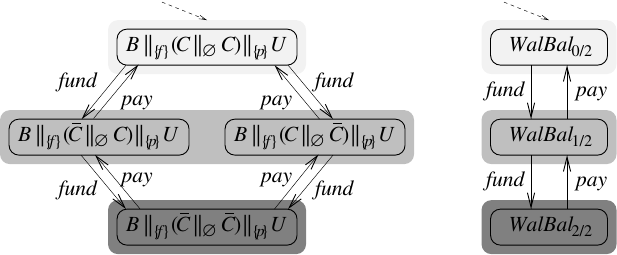}}
In the states of the automaton on the left-hand side, every process name and every action has been shortened
with its initial. Moreover, $\bar{C}$ stands for $\mathit{pay} \, . \, \mathit{Cell}$.

It turns out that $\mathit{WB}_{{\rm conc}, 2}$ is strongly bisimilar~\cite{Mil89a} to
$\mathit{WalBal}_{0/2}$, i.e., they are able to mimic each other's behavior stepwise. The bisimulation
proving this fact has been represented graphically by giving the same color to states in the same
equivalence class and different colors to different equivalence classes. The depicted relation is a strong
bisimulation because in both automata:

		\begin{itemize}

\item A light gray state can only reach a gray state by executing $\mathit{fund}$.

\item A gray state can only reach a dark gray state by executing $\mathit{fund}$ or a light gray state by
executing $\mathit{pay}$.

\item A dark gray state can only reach a gray state by executing $\mathit{pay}$.

		\end{itemize}

The second option is a pipeline implementation, in which the two-position wallet is obtained as the parallel
composition of two communicating one-position wallet cells:
\cws{0}{\begin{array}{rcl}
\mathit{WB}_{{\rm pipe}, 2} & \eqdef & \mathit{Bank} \, \pco{\{ \mathit{fund} \}} \, (\mathit{LCell} \,
\pco{\{ \mathit{move} \}} \, \mathit{RCell}) \, / \, \{ \mathit{move} \} \, \pco{\{ \mathit{pay} \}} \,
\mathit{User} \\
\mathit{Bank} & \eqdef & \mathit{fund} \, . \, \mathit{Bank} \\
\mathit{LCell} & \eqdef & \mathit{fund} \, . \, \mathit{move} \, . \, \mathit{LCell} \\
\mathit{RCell} & \eqdef & \mathit{move} \, . \, \mathit{pay} \, . \, \mathit{RCell} \\
\mathit{User} & \eqdef & \mathit{pay} \, . \, \mathit{User} \\
\end{array}}
Action $\mathit{move}$ models the passage of one CBDC unit from the left wallet cell to the right wallet
cell and occurs in the synchronization set of $\mathit{LCell} \, \pco{\{ \mathit{move} \}} \,
\mathit{RCell}$ as well as in the hiding set of $/ \, \{ \mathit{move} \}$ applied to the previous
subprocess. We have decided to hide the execution of $\mathit{move}$, which thus becomes the unobservable
action $\tau$ within transition labels, as it represents an implementation detail that should not be
perceived by an external observer. In a CBDC interpretation, this internal action can be viewed as a local
wallet operation that turns newly loaded value into immediately spendable offline~value.

To prove that this a correct implementation of the wallet balance management specification, similar to the
concurrent implementation we have to investigate the existence of some relation between $\mathit{WB}_{{\rm
pipe}, 2}$ and $\mathit{WalBal}_{0/2}$. Thus the first step consists of comparing the automata underlying
$\mathit{WB}_{{\rm pipe}, 2}$ and $\mathit{WalBal}_{0/2}$, which are shown below: \\[0.2cm]
\centerline{\includegraphics{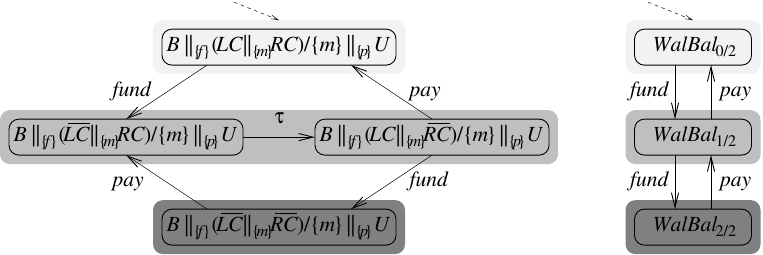}}
In addition to the same shorthands as before for process names and actions, on the left-hand side we have
$\overline{\mathit{LC}}$ for $\mathit{move} \, . \, \mathit{LCell}$ and $\overline{\mathit{RC}}$ for
$\mathit{pay} \, . \, \mathit{RCell}$.

It turns out that $\mathit{WB}_{{\rm pipe}, 2}$ is weakly bisimilar~\cite{Mil89a} to
$\mathit{WalBal}_{0/2}$, in the sense that $\tau$-actions are ignored in the bisimulation game. The weak
bisimulation proving this fact has been represented graphically by means of colors. The depicted relation is
a weak bisimulation because in both automata:

		\begin{itemize}

\item A light gray state can only reach a gray state by executing $\mathit{fund}$.

\item A gray state can only reach a dark gray state by executing $\mathit{fund}$ (possibly preceded by
$\tau$), a light gray state by executing $\mathit{pay}$ (possibly preceded by~$\tau$), or a gray state by
executing $\tau$ or staying idle.

\item A dark gray state can only reach a gray state by executing $\mathit{pay}$.

		\end{itemize}

We conclude by pointing out that, in the considered simplified scenario, it is easy to establish the
equivalence of both implementations to the specification. However, without the use of automated tools rooted
in formal methods, it would become very hard, if not impossible, to address realistic scenarios featuring
multiple banks and users that work in parallel on a large population of offline wallets, possibly including
wallet limits, synchronization with online ledgers, and anti-fraud mechanisms.

\section{Conclusions}
\label{sec:concl}

Contrary to what happens with SWIFT and other commercial bank circuits, CBDCs like the digital euro will be
deployed directly on the Internet. This yields the highest level of risk in terms of hacking and of rapidly
generating counterfeit money. Moreover, the need to provide an offline solution for digital euro tokens --
to offer cash-like features and promote financial inclusion -- constitutes, in itself, a problem in search
of a solution, especially in cases where some devices remain offline for a long time.

On the other hand, due to the impossibility results mentioned in Section~\ref{sec:sw_quality} -- concerning
the inability in general to theoretically guarantee the quality and correctness of the protocols and the
software required to implement them -- it is not possible to absolutely ensure the perfect and secure
functioning of a CBDC infrastructure. This leaves the risk of possible trapdoors, open to exploitation by
bad actors, which could be used to create counterfeit transactions and money.

The formal instruments briefly described in Section~\ref{sec:sw_quality} -- automata, process algebras,
logics, model checking, equivalence checking, deductive verification -- together with the related software
tools -- which have been successfully employed in large-scale case studies about formal modeling and
verification -- could help mitigate the aforementioned risks by enabling the automated validation of all
processes and protocols involved in a highly safety-critical financial system such as the one based on the
implementation of the digital euro. To the best of our knowledge, they have never been used for this purpose
before; therefore, we strongly recommend their adoption. The trust and reputation of the ECB are at stake,
with no room for errors.

\medskip
\noindent
\textbf{Acknowledgments.} This research has been supported by the PRIN 2020 project \linebreak \emph{NiRvAna
-- Noninterference and Reversibility Analysis in Private Blockchains}. The scholarship of the second author
at the Italian PhD Program in Blockchain and Distributed Ledger Technology has been funded by \emph{PNRR --
Piano Nazionale di Ripresa e Resilienza} according to D.M.~118/2023.

\bibliographystyle{plain}
\bibliography{biblio}

@inproceedings{And_1999,
Author = {R.J.~Anderson},
Title = {The Formal Verification of a Payment System},
Booktitle = {Industrial-Strength Formal Methods in Practice},
Publisher = {Springer},
Pages = {43--52},
Year = {1999}}

@misc{AC_2025,
Author = {{Atlantic Council}},
Title = {Central Bank Digital Currency Tracker},
Note = {\url{https://www.atlanticcouncil.org/cbdctracker/},
	last accessed on 21 July 2025},
Year = {2025}}

@techreport{BIS_2023_Polaris_1,
Author = {{Bank for International Settlements}\bibtrick{a}},
Title = {Project {P}olaris: A Handbook for Offline Payments with {CBDC}},
Note = {Part 1 of the Project Polaris series},
Month = {May},
Year = {2023}}

@techreport{BIS_2023_Polaris_2,
Author = {{Bank for International Settlements}\bibtrick{b}},
Title = {Project {P}olaris: A Security and Resilience Framework for {CBDC} Systems},
Note = {Part 2 of the Project Polaris series},
Month = {July},
Year = {2023}}

@techreport{BIS_2023_Polaris_3,
Author = {{Bank for International Settlements}\bibtrick{c}},
Title = {Project {P}olaris: Closing the {CBDC} Cyber Threat Modelling Gaps},
Note = {Part 3 of the Project Polaris series},
Month = {July},
Year = {2023}}

@techreport{BIS_2023_Polaris_4,
Author = {{Bank for International Settlements}\bibtrick{d}},
Title = {Project {P}olaris: A High-Level Design Guide for Offline Payments with {CBDC}},
Note = {Part 4 of the Project Polaris series},
Month = {October},
Year = {2023}}

@techreport{BoC_2023,
Author = {C.~Minwalla and J.~Miedema and S.~Hernandez and A.~Sutton-Lalani},
Title = {A Central Bank Digital Currency for Offline Payments},
Type = {Bank of Canada Staff Analytical Note},
Number = {2023-2},
Year = {2023}}

@techreport{BoE_2020,
Author = {{Bank of England}},
Title = {Central Bank Digital Currency: Opportunities, Challenges and Design},
Year = {2020}}

@techreport{BoF_2023,
Author = {J.~Nurminen and J.~Schreck},
Title = {Reining in the Expectations of Offline Payments},
Type = {Bank of Finland Bulletin series},
Number = {A:130},
Year = {2023}}

@techreport{BoS_2017_Ekrona_1,
Author = {{Sveriges Riksbank}},
Title = {The {R}iksbank's E-Krona Project: Report 1},
Year = {2017}}

@techreport{BoS_2018_Ekrona_2,
Author = {{Sveriges Riksbank}},
Title = {The {R}iksbank's E-Krona Project: Report 2},
Year = {2018}}

@techreport{BoS_2022_Ekrona_3,
Author = {{Sveriges Riksbank}},
Title = {E-Krona Pilot Phase 2},
Year = {2022}}

@techreport{BEFRG_2025,
Author = {A.~Esposito and F.P.~Rossi and M.~Bernardo and F.~Fabris and H.~Garavel},
Title = {Formal Modeling and Verification of the {A}lgorand Consensus Protocol in {CADP}},
Type = {ar{X}iv:2508.19452},
Year = {2025}}

@techreport{SS_2019,
Author = {S.~Sarswat and A.K.~Singh},
Title = {Formal Verification of Trading in Financial Markets},
Type = {ar{X}iv:1907.07885},
Year = {2019}}

@techreport{BZKWCC_2024,
Author = {C.~Beer and S.~Zingg and K.~Kostiainen and K.~W\"ust and V.~Capkun and S.~Capkun},
Title = {PayOff: A Regulated Central Bank Digital Currency with Private Offline Payments},
Type = {ar{X}iv:2408.06956},
Year = {2024}}

@techreport{Chainalysis_2025,
Author = {Chainalysis},
Title = {2025 Crypto Crime Report: Trends in Illicit Cryptocurrency Activity},
Year = {2025}}

@techreport{EUBOF_2021,
Author = {{EU Blockchain Observatory and Forum}},
Title = {Central Bank Digital Currencies and a Euro for the Future},
Year = {2021}}

@techreport{ECB_2020,
Author = {{European Central Bank}},
Title = {Report on a Digital Euro},
Year = {2020}}

@techreport{ECB_2024,
Author = {{European Central Bank}},
Title = {Economic Bulletin},
Type = {Number},
Number = {4},
Year = {2024}}

@misc{ECB_2025,
Author = {{European Central Bank}},
Title = {Digital Euro},
Note = {\url{https://www.ecb.europa.eu/euro/digital_euro/html/index.en.html},
	last accessed on 21 July 2025},
Year = {2025}}

@techreport{IMF_2024,
Author = {{International Monetary Fund}},
Title = {Cyber Risk: A Growing Concern for Macrofinancial Stability},
Type = {Global Financial Stability Report},
Number = {April 2024},
Year = {2024}}

@inproceedings{BZ_2018,
Author = {M.~Bartoletti and R.~Zunino},
Title = {{BitML}: A Calculus for {B}itcoin Smart Contracts},
Booktitle = {Proc.\ of the 25th ACM SIGSAC Conf.\ on Computer and Communications Security (CCS~2018)},
Publisher = {ACM Press},
Pages = {83--100},
Year = {2018}}

@inproceedings{LRK_2013,
Author = {C.~Lange and C.~Rowat and M.~Kerber},
Title = {The {ForMaRE} Project -- {F}ormal Mathematical Reasoning in Economics},
Booktitle = {Proc.\ of the 6th Int.\ Conf.\ on Intelligent Computer Mathematics (CICM~2013)},
Publisher = {Springer},
Series = {LNCS},
Volume = {7961},
Pages = {330--334},
Year = {2013}}

@inproceedings{MMNOSV_2018,
Author = {F.~Martinelli and F.~Mercaldo and V.~Nardone and A.~Orlando and A.~Santone and G.~Vaglini},
Title = {Safety Critical Systems Formal Verification Using Execution Traces},
Booktitle = {Proc.\ of the 27th IEEE Int.\ Conf.\ on Enabling Technologies: Infrastructure for Collaborative
	     Enterprises (WETICE~2018)},
Publisher = {IEEE-CS Press},
Pages = {247--250},
Year = {2018}}

@inproceedings{PI_2017,
Author = {G.O.~Passmore and D.~Ignatovich},
Title = {Formal Verification of Financial Algorithms},
Booktitle = {Proc.\ of the 26th Int.\ Conf.\ on Automated Deduction (CADE~2017)},
Publisher = {Springer},
Series = {LNAI},
Volume = {10395},
Pages = {26--41},
Year = {2017}}

@inproceedings{SIR_2013,
Author = {A.~Santone and V.~Intilangelo and D.~Raucci},
Title = {Efficient Formal Verification in Banking Processes},
Booktitle = {Proc.\ of the 9th IEEE World Congress on Services (SERVICES~2013)},
Publisher = {IEEE-CS Press},
Pages = {325--332},
Year = {2013}}

@book{Sto_2023,
Author = {J.H.~Stoel},
Title = {Solving the Bank: Lightweight Specification and Verification Techniques for Enterprise Software},
Publisher = {PhD Thesis},
Year = {2023}}

@article{VYC_2022,
Author = {S.~Verma and D.~Yadav and G.~Chandra},
Title = {Introduction of Formal Methods in Blockchain Consensus Mechanism and Its Associated Protocols},
Journal = {IEEE Access},
Volume = {10},
Pages = {66611--66624},
Year = {2022}}

@book{Ves_2023,
Author = {A.~Veschetti},
Title = {A Formal Analysis of Blockchain Consensus},
Publisher = {PhD Thesis},
Year = {2023}}

@techreport{WBG_2021,
Author = {{World Bank Group}},
Title = {The Global Findex Database 2021: Financial Inclusion, Digital Payments, and Resilience in the Age
	 of {COVID}-19},
Year = {2021}}

@book{AILS07,
Author = {L.~Aceto and A.~Ingolfsdottir and K.G.~Larsen and J.~Srba},
Title = {Reactive Systems: Modelling, Specification and Verification},
Publisher = {Cambridge University Press},
Year = {2007}}

@book{ABBHSU16,
Author = {W.~Ahrendt and B.~Beckert and R.~Bubel and R.~H\"{a}hnle and P.H.~Schmitt and M.~Ulbrich
	  (editors)},
Title = {Deductive Software Verification -- The KeY Book},
Publisher = {Springer},
Year = {2016}}

@book{ABCDF95,
Author = {M.~{Ajmone Marsan} and G.~Balbo and G.~Conte and S.~Donatelli and G.~Franceschinis},
Title = {Modelling with Generalized Stochastic {P}etri Nets},
Publisher = {John Wiley \& Sons},
Note = {\url{https://www.di.unito.it/~greatspn/index.html}},
Year = {1995}}

@book{ABC10,
Author = {A.~Aldini and M.~Bernardo and F.~Corradini},
Title = {A Process Algebraic Approach to Software Architecture Design},
Publisher = {Springer},
Note = {\url{http://www.sti.uniurb.it/bernardo/twotowers/}},
Year = {2010}}

@book{BW90,
Author = {J.C.M.~Baeten and W.P.~Weijland},
Title = {Process Algebra},
Publisher = {Cambridge University Press},
Year = {1990}}

@book{BK08,
Author = {C.~Baier and J.-P.~Katoen},
Title = {Principles of Model Checking},
Publisher = {MIT Press},
Year = {2008}}

@book{BPS01,
Author = {J.A.~Bergstra and A.~Ponse and S.A.~Smolka (editors)},
Title = {Handbook of Process Algebra},
Publisher = {Elsevier},
Year = {2001}}

@book{BC04,
Author = {Y.~Bertot and P.~Cast\'{e}ran},
Title = {Interactive Theorem Proving and Program Development},
Publisher = {Springer},
Note = {\url{https://rocq-prover.org/}},
Year = {2004}}

@book{BN19,
Author = {R.~Buyya and S.~Narayana Srirama (editors)},
Title = {Fog and Edge Computing: Principles and Paradigms},
Publisher = {Wiley},
Year = {2019}}

@article{Cho59,
Author = {N.~Chomsky},
Title = {On Certain Formal Properties of Grammars},
Journal = {Information and Control},
Volume = {2},
Pages = {137--167},
Year = {1959}}

@article{CCGR00,
Author = {A.~Cimatti and E.M.~Clarke and F.~Giunchiglia and M.~Roveri},
Title = {{NuSMV}: A New Symbolic Model Checker},
Journal = {Software Tools for Technology Transfer},
Volume = {2},
Pages = {410--425},
Note = {\url{https://nusmv.fbk.eu/}},
Year = {2000}}

@book{CGP99,
Author = {E.M.~Clarke and O.~Grumberg and D.A.~Peled},
Title = {Model Checking},
Publisher = {MIT Press},
Year = {1999}}

@article{Dij75,
Author = {E.W.~Dijkstra},
Title = {Guarded Commands, Nondeterminacy and Formal Derivation of Programs},
Journal = {Communications of the ACM},
Volume = {18},
Pages = {453--457},
Year = {1975}}

@book{DKV09,
Author = {M.~Droste and W.~Kuich and H.~Vogler (editors)},
Title = {Handbook of Weighted Automata},
Publisher = {Springer},
Year = {2009}}

@book{EB23,
Author = {T.~Erl and E.~Barcel\'o Monroy},
Title = {Cloud Computing: Concepts, Technology, Security, and Architecture},
Publisher = {Pearson},
Year = {2023}}

@article{EABR25,
Author = {A.~Esposito and A.~Aldini and M.~Bernardo and S.~Rossi},
Title = {Noninterference Analysis of Reversible Systems: An Approach Based on Branching Bisimilarity},
Journal = {Logical Methods in Computer Science},
Volume = {21(1)},
Pages = {6:1--6:28},
Year = {2025}}

@article{FLP85,
Author = {M.J.~Fischer and N.A.~Lynch and M.S.~Paterson},
Title = {Impossibility of Distributed Consensus with One Faulty Process},
Journal = {Journal of the ACM},
Volume = {32},
Pages = {374--382},
Year = {1985}}

@inproceedings{FG01,
Author = {R.~Focardi and R.~Gorrieri},
Title = {Classification of Security Properties},
Booktitle = {Proc.\ of the 1st Int.\ School on Foundations of Security Analysis and Design (FOSAD~2000)},
Publisher = {Springer},
Series = {LNCS},
Volume = {2171},
Pages = {331--396},
Year = {2001}}

@book{GHR94,
Author = {D.M.~Gabbay and I.~Hodkinson and M.~Reynolds},
Title = {Temporal Logic: Mathematical Foundations and Computational Aspects},
Publisher = {Oxford University Press},
Year = {1994}}

@article{GLMS13,
Author = {H.~Garavel and F.~Lang and R.~Mateescu and W.~Serve},
Title = {{CADP} 2011: A Tool for the Construction and Analysis of Distributed Processes},
Journal = {Software Tools for Technology Transfer},
Volume = {15},
Pages = {89--107},
Note = {\url{https://cadp.inria.fr/}},
Year = {2013}}

@inproceedings{GM82,
Author = {J.A.~Goguen and J.~Meseguer},
Title = {Security Policies and Security Models},
Booktitle = {Proc.\ of the 2nd IEEE Symp.\ on Security and Privacy (SSP~1982)},
Publisher = {IEEE-CS Press},
Pages = {11--20},
Year = {1982}}

@book{GM14,
Author = {J.F.~Groote and M.R.~Mousavi},
Title = {Modeling and Analysis of Communicating Systems},
Publisher = {MIT Press},
Note = {\url{https://www.mcrl2.org/}},
Year = {2014}}

@article{HHHK13,
Author = {E.M.~Hahn and A.~Hartmanns and H.~Hermanns and J.-P.~Katoen},
Title = {A Compositional Modelling and Analysis Framework for Stochastic Hybrid Systems},
Journal = {Formal Methods in System Design},
Volume = {43},
Pages = {191--232},
Note = {\url{https://www.modestchecker.net/}},
Year = {2013}}

@book{Han92,
Author = {H.~Hansson},
Title = {Time and Probability in Formal Design of Distributed Systems},
Publisher = {PhD Thesis},
Year = {1992}}

@book{Hen88a,
Author = {M.~Hennessy},
Title = {Algebraic Theory of Processes},
Publisher = {MIT Press},
Year = {1988}}

@book{Her02,
Author = {H.~Hermanns},
Title = {Interactive Markov Chains},
Publisher = {Springer},
Note = {Volume 2428 of LNCS},
Year = {2002}}

@book{Hil96,
Author = {J.~Hillston},
Title = {A Compositional Approach to Performance Modelling},
Publisher = {Cambridge University Press},
Note = {\url{https://www.dcs.ed.ac.uk/pepa/tools/}},
Year = {1996}}

@article{Hoa69,
Author = {C.A.R.~Hoare},
Title = {An Axiomatic Basis for Computer Programming},
Journal = {Communications of the ACM},
Volume = {12},
Pages = {576--580},
Year = {1969}}

@book{Hoa85,
Author = {C.A.R.~Hoare},
Title = {Communicating Sequential Processes},
Publisher = {Prentice Hall},
Year = {1985}}

@book{Hol03,
Author = {G.J.~Holzmann},
Title = {The {S}pin Model Checker: Primer and Reference Manual},
Publisher = {Addison-Wesley},
Note = {\url{https://spinroot.com/}},
Year = {2003}}

@book{HMU06,
Author = {J.E.~Hopcroft and R.~Motwani and J.D.~Ullman},
Title = {Introduction to Automata Theory, Languages, and Computation},
Publisher = {Pearson Addison-Wesley},
Year = {2006}}

@book{HC77,
Author = {G.E.~Hughes and M.J.~Creswell},
Title = {An Introduction to Modal Logic},
Publisher = {Methuen},
Year = {1977}}

@book{KKSS24,
Author = {U.~Kamath and K.~Keenan and G.~Somers and S.~Sorenson},
Title = {Large Language Models: A Deep Dive},
Publisher = {Springer},
Year = {2024}}

@misc{KN12,
Author = {S.~King and S.~Nadal},
Title = {{PPCoin}: Peer-to-Peer Crypto-Currency with Proof-of-Stake},
Note = {\url{https://peercoin.net/assets/paper/peercoin-paper.pdf}},
Year = {2012}}

@inproceedings{KNP11,
Author = {M.~Kwiatkowska and G.~Norman and D.~Parker},
Title = {{PRISM} 4.0: Verification of Probabilistic Real-Time Systems},
Booktitle = {Proc.\ of the 23rd Int.\ Conf.\ on Computer Aided Verification (CAV~2011)},
Publisher = {Springer},
Series = {LNCS},
Volume = {6806},
Pages = {585--591},
Note = {\url{https://www.prismmodelchecker.org/}},
Year = {2011}}

@book{LMD13,
Author = {P.~Lalanda and J.A.~McCann and A.~Diaconescu},
Title = {Autonomic Computing: Principles, Design and Implementation},
Publisher = {Springer},
Year = {2013}}

@article{Lan61,
Author = {R.~Landauer},
Title = {Irreversibility and Heat Generation in the Computing Process},
Journal = {IBM Journal of Research and Development},
Volume = {5},
Pages = {183--191},
Year = {1961}}

@article{LPY97,
Author = {K.G.~Larsen and P.~Petterson and Wang~Yi},
Title = {Uppall in a Nutshell},
Journal = {Software Tools for Technology Transfer},
Volume = {1},
Pages = {134--152},
Note = {\url{https://uppaal.org/}},
Year = {1997}}

@book{Mcm93,
Author = {K.L.~McMillan},
Title = {Symbolic Model Checking},
Publisher = {Springer},
Year = {1993}}

@book{Mil20,
Author = {M.~Milenkovic},
Title = {Internet of Things: Concepts and System Design},
Publisher = {Springer},
Year = {2020}}

@book{Mil89a,
Author = {R.~Milner},
Title = {Communication and Concurrency},
Publisher = {Prentice Hall},
Year = {1989}}

@book{Mul93,
Author = {S.~{Mullender (editor)}},
Title = {Distributed Systems},
Publisher = {Addison-Wesley},
Year = {1993}}

@misc{Nak08,
Author = {S.~Nakamoto},
Title = {Bitcoin: A Peer-to-Peer Electronic Cash System},
Note = {\url{https://bitcoin.org/bitcoin.pdf}},
Year = {2008}}

@book{NC10,
Author = {M.A.~Nielsen and I.L.~Chuang},
Title = {Quantum Computation and Quantum Information},
Publisher = {Cambridge University Press},
Year = {2010}}

@article{Ohe19,
Author = {P.~O'Hearn},
Title = {Separation Logic},
Journal = {Communications of the ACM},
Volume = {62},
Pages = {86--95},
Year = {2019}}

@article{Pau89,
Author = {L.C.~Paulson},
Title = {The Foundation of a Generic Theorem Prover},
Journal = {Journal of Automated Reasoning},
Volume = {5},
Pages = {363--397},
Note = {\url{https://www.cl.cam.ac.uk/research/hvg/Isabelle/}},
Year = {1989}}

@book{Pet62,
Author = {C.A.~Petri},
Title = {Kommunikation mit Automaten},
Publisher = {PhD Thesis},
Year = {1962}}

@book{Rei85,
Author = {W.~Reisig},
Title = {Petri Nets: An Introduction},
Publisher = {Springer-Verlag},
Year = {1985}}

@article{Ric53,
Author = {H.G.~Rice},
Title = {Classes of Recursively Enumerable Sets and Their Decision Problems},
Journal = {Trans.\ of the American Mathematical Society},
Volume = {74},
Pages = {358--366},
Year = {1953}}

@book{RN20,
Author = {S.~Russell and P.~Norvig},
Title = {Artificial Intelligence: A Modern Approach},
Publisher = {Pearson},
Year = {2020}}

@book{Seg95a,
Author = {R.~Segala},
Title = {Modeling and Verification of Randomized Distributed Real-Time Systems},
Publisher = {PhD Thesis},
Year = {1995}}

@techreport{Sza94,
Author = {N.~Szabo},
Title = {Smart Contracts},
Year = {1994}}

@article{Tur36,
Author = {A.M.~Turing},
Title = {On Computable Numbers, with an Application to the Entscheidungsproblem},
Journal = {Proc.\ of the London Mathematical Society},
Volume = {s2-42},
Pages = {230--265},
Year = {1936}}

@book{WBK20,
Author = {J.~Watt and R.~Borhani and A.K.~Katsaggelos},
Title = {Machine Learning Refined: Foundations, Algorithms, and Applications},
Publisher = {Cambridge University Press},
Year = {2020}}

@misc{WikiSwBugs,
Author = {Wikipedia},
Title = {List of Software Bugs},
Note = {\url{https://en.wikipedia.org/wiki/List_of_software_bugs}},
Year = {2008}}

\end{document}